\pdfoutput=1
\documentclass[12pt,amssymb]{article}

\usepackage{graphicx}
\usepackage{amsfonts}
\usepackage{amsmath}
\usepackage[usenames]{color}
\usepackage{amssymb}
\usepackage{amsthm}
\usepackage[mathscr]{eucal} 
\usepackage{url}

\usepackage[svgnames]{xcolor}
\usepackage[colorlinks,citecolor=DarkGreen,linkcolor=FireBrick,urlcolor=FireBrick,linktocpage,unicode]{hyperref}

\def\bZ{\mathbb{Z}}

\def\bR{\mathbb{R}}
\def\bC{\mathbb{C}}

\begin{document}

\baselineskip 0.6cm
\newcommand{\nequiv}{\mbox{\ooalign{\hfil/\hfil\crcr$\equiv$}}}
\newcommand{\nsupset}{\mbox{\ooalign{\hfil/\hfil\crcr$\supset$}}}
\newcommand{\nni}{\mbox{\ooalign{\hfil/\hfil\crcr$\ni$}}}
\newcommand{\nin}{\mbox{\ooalign{\hfil/\hfil\crcr$\in$}}}

\newcommand{\vev}[1]{ \left\langle {#1} \right\rangle }
\newcommand{\bra}[1]{ \langle {#1} | }
\newcommand{\ket}[1]{ | {#1} \rangle }
\newcommand{\Dsl}{\mbox{\ooalign{\hfil/\hfil\crcr$D$}}}
\newcommand{\Slash}[1]{{\ooalign{\hfil/\hfil\crcr$#1$}}}
\newcommand{\EV}{ {\rm eV} }
\newcommand{\KEV}{ {\rm keV} }
\newcommand{\MEV}{ {\rm MeV} }
\newcommand{\GEV}{ {\rm GeV} }
\newcommand{\TEV}{ {\rm TeV} }

\def\diag{\mathop{\rm diag}\nolimits}
\def\tr{\mathop{\rm tr}}

\def\Spin{\mathop{\rm Spin}}
\def\SO{\mathop{\rm SO}}
\def\O{\mathop{\rm O}}
\def\SU{\mathop{\rm SU}}
\def\U{\mathop{\rm U}}
\def\Sp{\mathop{\rm Sp}}
\def\SL{\mathop{\rm SL}}

\def\change#1#2{{\color{blue}#1}{\color{red} [#2]}\color{black}\hbox{}}

\def\Nequals#1{$\mathcal{N}{=}#1$}

\theoremstyle{definition}
\newtheorem{thm}{Theorem}[subsubsection]
\newtheorem{defn}[thm]{Definition}
\newtheorem{exmpl}[thm]{Example}
\newtheorem{props}[thm]{Proposition}
\newtheorem{lemma}[thm]{Lemma}
\newtheorem{rmk}[thm]{Remark}
\newtheorem{anythng}[thm]{}

\newenvironment{modified}{\bgroup}{\egroup}
\def\modi#1{#1}

 \begin{titlepage}
  
\begin{flushright}
 IPMU21-0070
\end{flushright}
 
  \vskip 3cm
  \begin{center}
  
  {\large \bfseries On odd number of fermion zero modes \\[.5em]
  on solitons in quantum field theory and string/M theory}
  
  \vskip 1.2cm
 
 Yotaro Sato, Yuji Tachikawa and Taizan Watari

 \vskip 0.4cm
  
  {\slshape
    Kavli Institute for the Physics and Mathematics of the Universe (WPI), \\
    the University of Tokyo, Kashiwa-no-ha 5-1-5, 277-8583, Japan
   }

 \end{center}
 \vskip 2.5cm
  
\noindent 
We argue that having an odd number of Majorana fermion zero modes on a 
dynamical point-like soliton signifies an inconsistency in a theory with 3+1 and higher dimensions.
We check this statement in a couple of examples in field theory and in string/M theory.

 \end{titlepage}

  \tableofcontents
 

\section{Introduction}

Fermion zero modes on solitons have played an important role 
in the study of quantum field theories and string/M theory.
For example, the spin and the flavor charge of an 't Hooft-Polyakov monopole 
in an $SU(2)$ gauge theory are determined 
by quantizing the fermion zero modes around it; 
its analysis goes back to Jackiw and Rebbi \cite{Jackiw:1975fn}.
To quantize the fermion zero modes on a soliton, we usually  regroup them into 
creation and annihilation operators, 
but this requires that the number of such zero modes should be even, 
when counted in a Majorana basis.
The quantization of $2n$ Majorana fermion zero modes leads to a Hilbert space of $2^n$ dimensions.

A more exotic possibility of having an odd number of Majorana fermion zero modes
was noticed and popularized by Kitaev \cite{Kitaev:2001kla} in 1+1 dimensions.
In this case, there is no way to quantize the fermion zero modes locally on a single soliton 
to assign a local Hilbert space.
For example, a single Majorana fermion would lead to a Hilbert space of dimension $\sqrt{2}$,
which is absurd.
Instead, what happens is that when there is a pair of such solitons, 
there is a Hilbert space $\bC^2$ associated to the pair, 
acted on by two Majorana fermion zero modes $\psi_1$ and $\psi_2$ 
localized at each of the solitons.
One natural question then is whether the same situation of 
having an odd number of Majorana fermion zero modes on a soliton is possible in higher dimensions.

In this short note, we argue that the answer is no in 3+1 dimensions and above, i.e.~%
that having an odd number of Majorana fermion zero modes 
on a dynamical point-like soliton in a theory with 3+1 dimensions or above
signifies an inconsistency.
We then examine this condition in a few examples, both in quantum field theory 
and in string/M theory.
We will see that this condition is satisfied in a nontrivial manner. 

\begin{modified}
Before proceeding, we note that the existence of a single Majorana fermion zero mode in 3+1 dimensions in a point-like configuration with a position-dependent mass term was originally discussed in \cite{Teo:2009qv}.
These configurations however carry additional orientational degrees of freedom
which allow nontrivial statistics to emerge, as studied in detail in \cite{Freedman:2010ak};
see also \cite{Freedman:2011zd}.
In this paper we restrict our attention to the cases where there are no such orientational degrees of freedom.\footnote{%
The authors thank J. McGreevy for the information contained in this paragraph.
}
\end{modified}

The rest of the paper is organized as follows. 
In Sec.~\ref{sec:argument} we provide the reasoning behind our consistency condition.
In Sec.~\ref{sec:QFT} we discuss quantum field theory examples, 
namely solitons in $SU(2)$ gauge theories with charged fermions;
the content of this section was already discussed in more detail in \cite{McGreevy:2011if}
and we only include a brief summary.
In Sec.~\ref{sec:M} we study a couple of examples in string/M theory,
by analyzing solitons arising from wrapped branes.
In Sec.~\ref{sec:conclusion} we conclude this paper by listing possible future directions.

\section{Majorana fermion zero modes and consistency}
\label{sec:argument}
\subsection{A consistent model in $1+1$ dimensions}
Let us start by recalling a model in  1+1 dimensions which has an odd number of Majorana fermion zero modes on a dynamical soliton.
We take a model with a real scalar $\phi$ and a non-chiral Majorana fermion $\psi$ in 1+1 dimensions.
We add a double-well potential $V(\phi)$ with two degenerate minima at $\phi=\pm \phi_0$,
and assume the coupling $m\phi \psi\psi$, where $m$ is a non-zero parameter.
In this model there is a dynamical domain wall connecting two vacua 
at $\phi=-\phi_0$ and $\phi=+\phi_0$.
Furthermore, it is well-known that there exists a single Majorana fermion zero mode on such a domain wall.

This makes it impossible to assign a local Hilbert space associated to a single domain wall.
This is due to the following:
A well-separated pair of a kink and an anti-kink has two nearly-degenerate ground states,
coming from the quantization of two Majorana fermion zero modes.
If a local Hilbert space $\mathcal{H}$ can be assigned to each of the solitons, 
its dimension has to satisfy $(\dim\mathcal{H})^2=2$, making $\dim\mathcal{H}=\sqrt2$, 
which is impossible.

When the scalar $\phi$ is made non-dynamical and the space is made discrete, 
the model reduces to Kitaev's quantum wire \cite{Kitaev:2001kla}.
We also note that in \cite{Jackiw:1975fn} Jackiw and Rebbi already studied 
a closely related model where $\psi$ is taken to be a complex Dirac fermion of charge $\pm1$ 
rather than a real Majorana fermion.
In this case, there is a complex conjugate pair of fermion zero modes on a kink,
whose quantization assigns the charge $\pm1/2$ to the soliton.

\subsection{A consistent model in $2+1$ dimensions}
Let us next recall a consistent model in $2+1$ dimensions,
 where a Majorana fermion zero mode arises on a dynamical soliton.
We simply consider a $U(1)$ gauge theory with a Dirac fermion $\psi$ of charge $+1$ 
and a complex scalar $\phi$ of charge $-2$, with the interaction $\phi\psi\psi + c.c.$.
We further introduce a potential $V(\phi)$ so that it has a minimum at $|\phi|\neq 0$.
The $U(1)$ gauge symmetry is broken to $\bZ_2$, and there are vortex solutions.
Furthermore,  it is known that there is a single Majorana fermion zero mode on the vortex
with the minimal winding number. 

This is a classic result going back to \cite{Jackiw:1981ee,Weinberg:1981eu},
whose importance in the physics of topological superconductors was recognized more recently.
A nice summary can be found in Sec.~3 and in Appendix B of \cite{Seiberg:2016rsg}.
As discussed there, the presence of an odd number of Majorana fermion zero modes
leads to the non-Abelian statistics possessed by the vortices.\footnote{%
This observation goes back to \cite{Moore:1991ks}.
}

\subsection{Inconsistency in $3+1$ dimensions and above}
\label{sec:X}

Let us now present an argument that one can never have an odd number of Majorana fermion zero modes on a point-like dynamical soliton in a consistent quantum theory in $d+1$ dimensions, when $d\ge 3$.
Suppose, on the contrary, that we have a soliton with an odd number, say $n$, of Majorana fermion zero modes, which we denote by $\psi_{1,\ldots,n}$.
It is impossible to assign a local Hilbert space associated to the soliton. 
We can still assign a Hilbert space to a pair of such solitons.
As there are $2n$ Majorana fermion zero modes in total, this Hilbert space has dimension $2^n$. Up to this point, there is no difference whether $d=1,2$ or $d\geq3$.

We now consider semi-classical quantization of this pair of solitons.
When we hold them at a large distance $L$ from each other,
the configuration space has the topology $S^{d-1}/\bZ_2$,
where the quotient comes from our assumption that we consider two identical solitons.
Recall that we assumed $d\ge 3$. 
Then  $\pi_1(S^{d-1}/\bZ_2)=\bZ_2$,
and the wavefunction can be thought of as a wavefunction on $S^{d-1}$ invariant under a $\bZ_2$ operation $P$.
As is well known, this was why there is the boson-fermion dichotomy\footnote{%
Anyons are allowed in $d=2$ because the representation of $\pi_1(S^{d-1})=\bZ$ is parameterized by a spin $\in U(1)$.
In $d=1$ we do not have a concept of spin, since $\pi_1(S^{d-1})$ is trivial.
} when $d\ge 3$,
depending on whether $P$ is realized as $+1$ or $-1$.

Let us first consider the case $n=1$ for simplicity.
The Majorana fermion zero modes  $\psi^{(1,2)}$ 
supported at each soliton act on $\bC^2$, as $\sigma_x$ and $\sigma_y$, say.
We then need an order-2 operation $P$ satisfying 
\begin{equation}
P\psi^{(1)} P^{-1}= \psi^{(2)},\qquad
P\psi^{(2)} P^{-1}= \psi^{(1)}. \label{bar}
\end{equation}
But this is clearly impossible,\footnote{%
Note that such a $P$ needs to exchange $\sigma_x$ and $\sigma_y$, 
and therefore is a $180^\circ$ rotation around the line $x=y$, $z=0$.
This squares to $-1$ when acting on a spinor.
See also Sec.~3.4 of \cite{Seiberg:2016rsg} for the review of exchange statistics of vortices, 
each carrying an odd number of Majorana fermion zero modes.}
since any such $P$ has  eigenvalues  $\pm i$ and therefore is of order 4.
This is against our requirement\footnote{%
Another more geometrical way to state the issue is the following.
We have a bundle of Clifford algebras generated by $\psi^{(1,2)}$ over the configuration space $S^{d-1}/\bZ_2$.
We then ask whether there is a bundle of two-dimensional representations of these Clifford algebras over $S^{d-1}/\bZ_2$.
What we showed here is that there is no such bundle of representations when $d\ge 3$.
} that $P$ is of order 2.

More generally, such $P$ exchanging two sets of $n$ Majorana fermion zero modes has eigenvalues $(\pm i)^n$, 
and therefore it is of order 4 when $n$ is odd.
This means that we cannot consistently form the wavefunction of a system of two identical solitons, each having an odd number of Majorana fermion zero modes,
if the spacetime has dimension $3+1$ or higher.

\section{Field theory examples}
\label{sec:QFT}

In this section we first study an example in $3+1$ dimensions,
and then discuss an analogue in  $4+1$ dimensions.
Both cases involve $SU(2)$ gauge theories with fermions in the doublet representation.
\begin{modified}
The content of this section was discussed more thoroughly by McGreevy and Swingle in \cite{McGreevy:2011if};
we only include a brief summary.
\end{modified}

\subsection{An example in $3+1$ dimensions}
\label{sec:Y}
We consider an $SU(2)$ gauge theory in 3+1 dimensions coupled to a scalar $\phi$ in the triplet,
and introduce $n$ Weyl fermions $\psi_{1,\ldots,n}$ in the doublet
with the coupling $\phi \psi \psi$, where the color indices are contracted appropriately. 

We give a vacuum expectation value to $\phi$ so that we have 't Hooft-Polyakov monopoles.
As was already computed in \cite{Jackiw:1975fn}, 
there is a single Majorana fermion zero mode localized at the monopole for each Weyl fermion in the doublet.
In total, there are $n$ Majorana fermion zero modes.
Therefore, the system becomes inconsistent when $n$ is odd, due to our condition in Sec. \ref{sec:X}.

Of course this is in agreement with the well-known fact 
that there is the $SU(2)$ global anomaly of Witten associated to $\pi_4(SU(2))=\bZ_2$
when $n$ is odd \cite{Witten:1982fp}.
Our analysis shows how the same anomaly can manifest in a rather different manner.
This point was discussed in detail in \cite{McGreevy:2011if}.

Note that this example also shows that it is perfectly possible to have an odd number of
Majorana fermion zero modes on \emph{static, external} point-like solitons.
Indeed, we can simply consider the same model where only the fermion field is considered dynamical,
while the $SU(2)$ gauge field and the scalar in the triplet are considered as background fields.
We can still introduce an 't Hooft-Polyakov monopole background in such a theory,
which would have an odd number of Majorana fermion zero modes when $n$ is odd.

\begin{modified}
It is also of interest to consider the model with the same field content 
where we make the triplet scalar field dynamical 
but keep $SU(2)$ symmetry non-dynamical. 
In this case, the solitons are dynamical, 
but the hedgehog scalar field configuration carries additional orientational degrees of freedom.
Then the configuration space of the two-soliton system is no longer $S^2/\bZ_2$ and is more complicated.
This  allows a more complicated non-Abelian statistics which comes with an odd number of Majorana fermionic zero modes per soliton. 
For more details, see \cite{Teo:2009qv,Freedman:2010ak}.
\end{modified}

Before proceeding, let us make a small digression, which might be of some interest to 
those familiar with the Seiberg-Witten theory.
Let us pretend that we did not know Witten's $SU(2)$ anomaly,
and try to analyze 4d \Nequals{2} $SU(2)$ gauge theory with $n$ half-hypermultiplets in the doublet.
We analyze the monodromy matrix $M$ associated to a loop around $u\sim \infty$ in the $u$-plane,
which is determined by the one-loop beta function thanks to the holomorphy.
The computation for even $n$ was given in the classic paper \cite{Seiberg:1994aj},
which can be carried over to the more general case.
The result is 
\begin{equation}
M=\begin{pmatrix}
-1 & 0\\
4-n/2 & -1
\end{pmatrix},
\end{equation}
which fails to be integer-valued when $n$ is odd. 
This inconsistency can be traced back to the problem of assigning electric charges to dyons
when there is an odd number of Majorana fermion zero modes on a monopole.\footnote{%
For a careful and pedagogical introduction to the effect of fermion zero modes on monopoles
in both non-supersymmetric and supersymmetric contexts, see \cite{Harvey:1996ur}.
}

\subsection{An example in $4+1$ dimensions}
\label{sec:Z}
Let us next consider an analogue in  $4+1$ dimensions.
Namely, let us take an $SU(2)$ gauge theory
with $n$ fermions $\psi$ in the doublet with the symplectic-Majorana condition.
Here, the symplectic-Majorana condition on a fermion field $\psi_{a\alpha}$ in the doublet
is imposed by \begin{equation}
\psi_{a\alpha}= \epsilon_{ab}J_{\alpha\beta}(\psi^*)^{b\beta}
\end{equation}
where $a,b=1,2$ are the $SU(2)$ indices,
$\alpha,\beta=1,2,3,4$ are the spinor indices,
and $\epsilon_{ab}$ and $J_{\alpha\beta}$ are the antisymmetric invariant tensors
which exist because $\mathbf{2}$ of $SU(2)$ and $\mathbf{4}$ of $SO(4,1)$ are pseudo-real.

We now note that an $SU(2)$ instanton configuration on $\bR^4$ can be regarded 
as a point-like soliton in a $(4+1)$-dimensional theory.
In this background, the index theorem tells us that there are $n$ Majorana fermion zero modes.
When $n$ is odd, there is an inconsistency discussed in Sec.~\ref{sec:X}.
At the same time, the theory is afflicted with a global anomaly, 
this time associated with $\pi_5(SU(2))=\bZ_2$ instead of $\pi_4(SU(2))=\bZ_2$.
Therefore, we again find that the number of Majorana fermion zero modes is even in a consistent theory.

To connect more directly with the discussions in Sec.~\ref{sec:Y},
we can add a triplet scalar field $\phi$ and give it a vacuum expectation value,
breaking $SU(2)$ to $U(1)$.
The 't Hooft-Polyakov monopole solution now gives rise to a string-like soliton,
on which there are $n$ Majorana-Weyl fermion zero modes.
We can now compactify the entire setup on $S^1$ with a periodic spin structure,
around which we wrap the string-like soliton.
We now find a point-like soliton with $n$ Majorana fermion zero modes.
Again we find that $n$ needs to be even.

\section{String/M theory examples}
\label{sec:M}

Let us next study solitons in compactifications of string/M theory.
In this section, we only discuss the case of point-like solitons
arising from compactifying a string-like soliton in higher dimensions,
as we saw at the end of Sec.~\ref{sec:Z}.

The condition that there cannot be an odd number of Majorana fermion zero modes on a point-like soliton in $ d+1$ dimensions
then translates to the condition that there cannot be an odd number of Majorana-Weyl fermions on a string-like soliton in $(d+1)+1$ dimensions.
For simplicity, we assume that the worldsheet theory is almost free 
at some scale along the renormalization group flow, with scalars and fermions.
Then, the number of Majorana-Weyl fermions is even or odd depending on whether
the difference of central charges  $c_L-c_R$ is an integer or an integer plus $1/2$.

Let us study our condition in the case of a single M5-brane wrapped on 4-cycles in spin manifolds.
The number of Majorana-Weyl fermions on the resulting string-like solitons
can be counted by explicitly counting the zero modes,
as was originally considered by Maldacena, Strominger and Witten in \cite{Maldacena:1997de}
when the manifold was further assumed to be Calabi-Yau.
Here we adopt a quicker approach of integrating the anomaly polynomial over the 4-cycle. 

The anomaly polynomial of a single M5-brane is \cite{Witten:1996hc,Monnier:2013rpa}
\begin{align}
    \label{6dAnomPoly}
    I_8=\frac{1}{48}\left[p_2(NW)-p_2(TW)+\frac{1}{4}(p_1(TW)-p_1(NW))^2\right]+\frac12\iota^*(G)^2,
\end{align} 
where $W$ is the worldvolume of the M5-brane; $N$ and $T$ are for the normal and tangent bundles, 
$G$ is the background 4-form flux of the spacetime,
and $\iota^*$ denotes the pull-back to the worldvolume.

Let us now integrate it over the 4-cycle $P$ within a 6-manifold $M$.
The 6d worldvolume theory reduces to the 2d worldsheet theory.
Denoting the Chern roots to $TP$ by $\pm\lambda_1$, $\pm\lambda_2$
and those to $NW$ by $\pm n_1$, $\pm n_2$, $0$,
we have
\begin{align}
    I_8=\frac{1}{48}\left[n_1^2n_2^2-\lambda_1^2\lambda_2^2-(\lambda_1^2+\lambda_2^2)p_1(T\Sigma)+\frac{1}{4}(\lambda_1^2+\lambda_2^2+p_1(T\Sigma)-n_1^2-n_2^2)^2\right]+\frac12\iota^*(G)^2.
\end{align}
Here, $T\Sigma$ is the tangent bundle of the worldsheet. Therefore, the anomaly polynomial of the worldsheet theory is given by 
\begin{align}
    I_4
    &=-\frac{p_1(T\Sigma)}{96}\int_P\left[\lambda_1^2+\lambda_2^2+n_1^2+n_2^2\right].
\end{align}
Here $G^2$ did not contribute, since we assume that the flux is only along $M$.

Recalling that $c_L-c_R$ and the anomaly polynomial of a worldsheet theory are related as \begin{equation}
I_4 = \frac{c_L-c_R}{24}p_1(T\Sigma)
\end{equation} in general, we see that \begin{align}
    \label{cGB}
    c_L-c_R=-\frac{1}{4}\int_P(\lambda_1^2+\lambda_2^2+n_1^2+n_2^2)=-\frac{1}{4}\int_P\iota^*p_1(TM).
\end{align}
When $M$ is a Calabi-Yau 3-fold, 
$c_L-c_R=\frac{1}{2}\int_P\iota^*c_2(TM)$ for an M5-brane because of the relation $p_1=c_1^2-2c_2$ on a complex manifold and the Calabi-Yau condition 
$c_1(TM)=0$. 
This can then be compared with the $c_L$, $c_R$ found in \cite{Witten:1996hc} and we find a nice agreement.

Our question is whether this $c_L-c_R$ given in \eqref{cGB} is an integer. 
For a spin manifold $M$, $p_1(TM)/2$ is known to be an integral class\footnote{%
This is because $p_1=w_2^2$ mod 2 for any orthogonal bundles,
and $w_2$ of a spin bundle is zero.
}.
Therefore, $c_L-c_R$ given in \eqref{cGB} is at worst a half-integer.
To see that this is actually an integer, we use two facts about M-theory.
The first is the shifted quantization law of the $G$-flux, 
originally found in \cite{Witten:1996md}: \begin{equation}
[G] - [\frac{p_1(TM)}4] \in H^4(TM,\bZ)
\end{equation}  
and the second is the Bianchi identity \cite{Howe:1996yn,Howe:1997fb,Chu:1997iw}
\begin{equation}
 -dH=\iota^*G
\end{equation} 
on the worldvolume of the M5-brane, where $H$ is the flux of the self-dual 2-form defined globally on it.
These two conditions mean that \begin{equation}
[\frac{\iota^*p_1(TM)}4] \in H^4(TP,\bZ), 
\end{equation}
which was what we wanted to show.\footnote{%
We note that when $M$ is simply-connected, spin and has no torsion in cohomology, 
Wall's theorem \cite{Wall}
says that $\int_M(4P^3-P\cdot p_1(TM))\in 24\bZ$.
This then means that $p_1(TM)/4$ is an integral class when pulled back to $P$.
The argument in the main text is applicable  more generally, in that $M$ may not be necessarily simply-connected,  may have torsion, and may be of any dimensions.
}

Before proceeding, let us make a digression on the relation to the $SU(2)$ gauge theory discussed in Sec.~\ref{sec:QFT}.
Although string/M theory is usually thought to be consistent,
it is difficult to show that it is actually the case in every case imaginable.
For example, establishing that no string/M theory construction gives 
a 4d $SU(2)$ gauge theory afflicted with Witten's $SU(2)$ anomaly has been a difficult problem.

This question was studied in the string/M-theory compactifications to four dimensions with \Nequals2 supersymmetry in \cite{Enoki:2020wel}.
It was also studied in the context of general heterotic compactifications in \cite{Tachikawa:2021mby}, 
which depended on a well-motivated but unproved mathematical conjecture.

Our analysis in this section can be thought of as providing another indirect piece of evidence that Witten's anomaly does not arise in the type IIA frame,
where the $SU(2)$ gauge group arises from the $\bC^2/\bZ_2$ singularity.
In such cases, the spontaneous breaking of symmetry to $U(1)$ would be given by a resolution of the singularity, making the internal manifold smooth.
Then the 't Hooft-Polyakov monopole would be given by a wrapped D4-brane.
Lifting the IIA setup to the M-theory, we find that 
the question of the number of Majorana fermion zero modes on 't Hooft-Polyakov monopoles
is mapped to the question of the number of Majorana fermion zero modes on M5-branes wrapped on 4-cycles, further compactified on $S^1$.

\section{Conclusions}
\label{sec:conclusion}

In this paper we studied the effect of having an odd number of Majorana fermion zero modes on dynamical point-like solitons.
We saw that it is perfectly consistent in $1+1$ and $2+1$ dimensional models,
while we argued that it signals an inconsistency if the models are in $3+1$ dimensions or higher.

As an explicit field-theoretical example, \modi{following \cite{McGreevy:2011if}},
we considered the 4d $SU(2)$  gauge theory with $n$ Weyl fermions in the doublet representation.
There, we have $n$ Majorana fermion zero modes on the 't Hooft-Polyakov monopole,
rendering the system inconsistent when $n$ is odd.
This gives the same consistency condition as the one from Witten's $SU(2)$ anomaly.

We then studied the point-like solitons in  string or M-theoretic constructions,
which arise from compactifications of string-like solitons of higher dimensional theories.
Our concrete example was the case of  M5-branes wrapped on 4-cycles in spin-manifolds.
The number of Majorana fermion zero modes always turned out to be even,
thanks to various previously-known consistency conditions of M-theory backgrounds.
We note that our M-theoretic analysis predicts an even number of zero modes for compactifications
down to 2+1 dimensions or less, for which our general argument allowed an odd number.

One possible future direction is to study many other explicit constructions in string/M theory
in order to check that there is always an even number of Majorana fermion zero modes.
In this paper we only studied the cases where the point-like solitons in question 
come from the $S^1$ compactification of string-like solitons.
There is in general no guarantee that point-like solitons in string/M theory is given in this manner.
In those cases, we would need to use mod-2 index theorems on the worldvolume of the branes to analyze the number of fermionic zero modes.

Another possible future direction is to give a better general argument for our condition than the one given in Sec.~\ref{sec:X}.
As is the case for any general no-go theorems in quantum field theory, 
we might have used some implicit assumptions,
and there might still be some loophole.
Although unlikely, it would be worthwhile to look for such a loophole,
since point-like solitons with an odd number of Majorana fermion zero modes 
would lead to particles whose statistics is neither fermionic nor bosonic.

 \subsection*{Acknowledgments}   

The authors thank Ryohei Kobayashi and Yunqin Zheng for discussions,
\modi{and John McGreevy for the references \cite{Teo:2009qv,Freedman:2010ak,Freedman:2011zd,McGreevy:2011if}}.
 This work is supported in part by 
JSPS Fellowship for Young Scientists (YS), the IGPEES program (YS), 
JSPS KAKENHI Grant-in-Aid (Kiban-S) No.16H06335 (YT),
WPI Initiative (all) 
and a Grant-in-Aid for Scientific Research on Innovative Areas 6003 (TW), 
MEXT, Japan.

\def\arxivfont{\rm}
\baselineskip=.94\baselineskip
\let\raggedright\relax
\bibliographystyle{ytphys}
\bibliography{ref}
\end{document}